\pdfoutput=1
\documentclass[%
 reprint,
superscriptaddress,
 amsmath,amssymb,
 prl,
floatfix,
]{revtex4-2}

\usepackage{graphicx}
\usepackage{lipsum}
\usepackage{dcolumn}
\usepackage{bm}
\usepackage{gensymb} 
\usepackage{xcolor}
\usepackage{amsmath}
\usepackage{physics}
\usepackage{bbold}
\definecolor{red}{rgb}{1,0,0}

\usepackage{hyperref}
\hypersetup{
colorlinks=true,
linkcolor=black,
filecolor=black,  
citecolor=black,
urlcolor=black,
}

\begin{document}


\title{Time reflection and refraction 
in synthetic frequency dimension}

\author{Olivia Y. Long}
\affiliation{Department of Applied Physics, Stanford University, Stanford, California 94305, USA}
\author{Kai Wang}
\affiliation{Ginzton Laboratory and Department of Electrical Engineering,
Stanford University, Stanford, California 94305, USA}

\author{Avik Dutt}
\affiliation{Department of Mechanical Engineering, University of Maryland, College Park, Maryland 20742, USA}

\author{Shanhui Fan}
\email{shanhui@stanford.edu}
\affiliation{Department of Applied Physics, Stanford University, Stanford, California 94305, USA}
\affiliation{Ginzton Laboratory and Department of Electrical Engineering,
Stanford University, Stanford, California 94305, USA}




\date{\today}

\begin{abstract}

The duality of space and time in Maxwell's equations has prompted interest in time boundaries and the accompanying temporal analog of spatial reflection and refraction. 
However, achieving observable time boundary effects at optical frequencies in real materials is challenging. In this work, we demonstrate that time reflection and refraction can be observed in a two-band model centered around a non-zero reference energy. Our model can be physically implemented in the synthetic frequency dimension as a system of two coupled dynamically-modulated ring resonators.
We find that modulation at microwave frequencies is sufficient to observe time boundary effects for optical waves in synthetic frequency dimension. Our work shows that implementing multi-band models in synthetic dimensions opens a new avenue for further exploration of time boundaries. 
\end{abstract}

\maketitle



 
The space-time duality inherent in Maxwell's equations suggests that there are temporal analogs to many of the more familiar spatial phenomena \cite{space-time_duality_temporal_imaging_kolner_1994, akhmanov_1969_space-time_analogy}.
Such parallels between space and time include the laws of reflection and refraction. Similar to a spatial boundary, a temporal boundary induces scattering of incident electromagnetic waves.
Previous theoretical works have demonstrated that at a temporal boundary created by changing the refractive index, time-refracted and time-reflected waves are generated 
\cite{reflec_trans_EM_waves_temp_boundary_agrawal_OL_2014, time_refract_reflection_Mendonca_2002, morgenthaler_velocity_mod_EM_waves_1958, temp_analog_reflec_refrac_optical_agrawal_PRL_2015}.  
 Such time boundaries have been further explored to form photonic time crystals 
 \cite{topolog_photonic_time_crystal_segev_optica_2018, Cervantes_PRA_2009, light_emission_free_elec_PTC_PNAS_2022, amplified_emission_lasing_PTC_segev_science_2022} and to achieve temporal aiming \cite{temporal_aiming_engheta_2020}, inverse prisms \cite{inverse_prism_OL_2018}, and higher-order transfer functions \cite{temp_multilayer_high_order_transfer_Alu_APL_2021}.











However, to observe time boundary effects experimentally for optical waves, the change in refractive index must be comparable to that of the material and occur on the time scale of the optical wave period (few femtoseconds)
\cite{exp_observ_fs_time_bound_segev_CLEO_2021, amplified_emission_lasing_PTC_segev_science_2022, boyd_time_refrac_ENZ_nature_2020}. 
%
%
%
These requirements are difficult to satisfy in real materials. 
To date, time refraction has been experimentally achieved at optical frequencies by using epsilon-near-zero media \cite{boyd_time_refrac_ENZ_nature_2020}. However, time reflection has only been observed in water waves \cite{time_reversal_holography_fink_nature_physics_2016}. 
%

 In this Letter, we propose using a two-band model to achieve time reflection and refraction.
 In this model, the time boundary is implemented by changing the coupling constants in time. We show that the required time scale for the modulation is no longer controlled by the frequency of the waves themselves, but rather by the band gap in the two-band model. Consequently, 
  time reflection and refraction effects can be observed in optical waves by applying a time boundary occurring on the microwave time scale, in contrast to previous works.



Furthermore, we  show that our system can be implemented naturally in synthetic dimensions. The concept of synthetic dimension allows one to transcend limits of spatial dimension by coupling different physical states to explore higher-dimensional systems.
In photonics, realizations of  synthetic dimensions include the use of orbital angular momentum, pulse arrival time, frequency modes, spatial supermodes, and mode propagation direction \cite{topolog_quantum_matter_synth_dim_ozawa_price_2019, luqi_synth_dim_photonics_Optica_2018, avik_HOTI_SSH_2ring_resonator_2020, single_photonic_cavity_2_synth_dims_avik_science_2020, high-dim_freq_crystals_loncar_optica_2020, quantum_sim_2d_topolog_physics_ZWZhou_nature_2015, chalabi_synth_gauge_field_2D_time_quantum_walk_PRL_2019, topolog_freq_conversion_quantum_PRX_2017, boundaries_with_memory_floquet_synthetic_crystals_PRL_2018, single_photonic_cavity_2_synth_dims_avik_science_2020, photonic_topolog_insulator_synth_dim_eran_nature_2019, boundaries_with_memory_floquet_synthetic_crystals_PRL_2018,  luqi_photonic_gauge_potential_synth_dim_OL_2016, synth_space_arbitrary_dim_ring_res_luqi_PRB_2018, creating_bound_synth_freq_avik_nature_2022, photonic_quantum_comp_synthetic_time_ben_optica_2021, siddarth_arbitrary_linear_trans_synth_dim_nature_2021}.
In this work, we propose to implement the two-band model in a system of coupled ring resonators to observe time reflection and refraction in synthetic frequency space. 



\begin{figure*}[t]
\includegraphics[width=\textwidth]{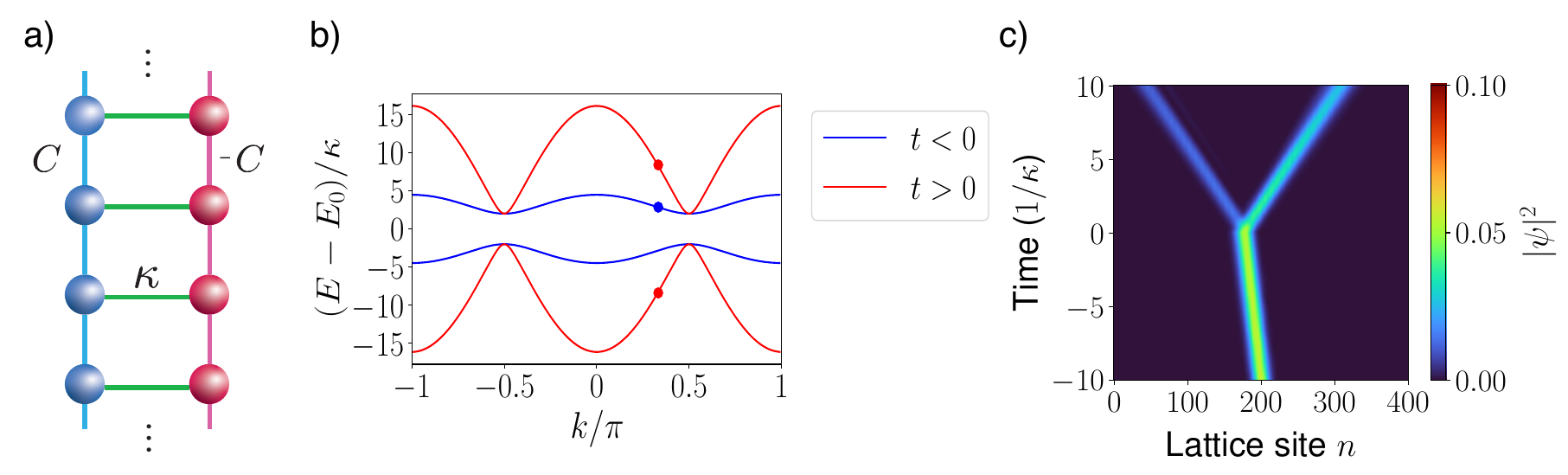}
\caption{\label{fig1} (a) Schematic of two-leg ladder lattice model. Blue and red spheres represent $a$ and $b$ lattice sites, respectively. $C$ and $-C$ are the inter-rung coupling constants for the $a$ and $b$ sublattices, respectively. $\kappa$ is the inter-leg coupling constant. (b) Band structure of Hamiltonian given by Eqn. \ref{hamilt}, where blue indicates before the time boundary ($t < 0$) with coupling constant $C/\kappa = 1$, while red indicates after the time boundary ($t > 0$) with $C/\kappa = -4$. Dots denote the states on the band structures with central wavevector $k_0 = \pi/3$. (c) Temporal evolution of the system initialized in the state described by Eqn. \ref{gaussian_eigenstate}. The system consists of $N=400$ total lattice sites, comprised of $M=200$ unit cells of $a$ and $b$ sites. The even and odd lattice sites are $a$ and $b$ sites, respectively. 
The coupling constant $C$ changes instantaneously from $C/\kappa = 1$ to $C/\kappa = -4$ at the time boundary $t=0$. Simulation 
employs the matrix exponential method.} 
\end{figure*}


 





We start by reviewing the basic working principles of a time boundary in a medium with dispersion $E(k)$. A time boundary can be implemented by a temporally sudden, but spatially uniform change of the medium's permittivity. For a wave with momentum $k$ and energy $E(k)$, such a time boundary will excite both the wave at $\widetilde{E}(k)$ and its time-reversed counterpart at  $-\widetilde{E}(k)$, where $\widetilde{E}(k)$ is the energy after the time boundary \cite{galiffi_photonics_time-varying_media_review_2022, time_refract_reflection_Mendonca_2002}. 
Importantly, to achieve such excitation, the temporal change of the permittivity must be fast enough to be comparable with the energy difference $2\widetilde{E}(k)$. 
For $\widetilde{E}(k)$ in the optical frequencies, it becomes very difficult to achieve such energy transitions, as the required switching speed is too fast.

In contrast to previous works, our approach is based on a periodic system with two energy bands that satisfy $E_+(k)+E_-(k)=2E_0$, where $E_+(k)$, $E_-(k)$ denote the upper and lower bands at momentum $k$, respectively, and $E_0$ is a constant reference energy. 
We will show that time reflection and refraction can be achieved by a sudden change of the energy bands, which results in the simultaneous excitation of the states in the two bands. In our case, the change of the band structure only needs to be fast enough as compared with the separation of the two energy bands. 
While $E_0$ can be at optical frequencies, it is possible to have $E_+(k) - E_-(k) \ll E_0 $. Thus, the required modulation speed is significantly lower than that of the previous approach.

We illustrate our approach by using a two-legged ladder model (Fig. \ref{fig1}a), as described by:
\begin{align} \label{hamilt}
    H = \sum_m E_0 [\hat{a}_m^\dagger \hat{a}_m + \hat{b}_m^\dagger \hat{b}_m] &+ \sum_m [C \hat{a}_m^\dagger \hat{a}_{m+1} + \text{h.c.}] \\
    + \sum_m [-C \hat{b}_m^\dagger \hat{b}_{m+1} + \text{h.c.}] \nonumber 
    &+ \sum_m [\kappa \hat{a}_m^\dagger \hat{b}_m + \text{h.c.}] 
\end{align}
In Eqn. \ref{hamilt}, $E_0$ is the on-site energy of each lattice site and h.c. stands for Hermitian conjugate. $\hat{a}_m^\dagger$, $\hat{b}_m^\dagger$ ($\hat{a}_m$, $\hat{b}_m$) are the creation (annihilation) operators for $a$ and $b$ lattice sites in the $m$th unit cell, respectively. $C$ is the coupling constant between nearest-neighbor $a$ sites. The coupling constant between the nearest neighbor $b$ sites are fixed at $-C$.
 The $a$ and $b$ sublattices thus each form a leg of the ladder model. $\kappa$ is the inter-leg coupling constant between $a$ and $b$ sites within a unit cell (Fig. \ref{fig1}a). In our system, the parameters $C$ and  $\kappa$ are real. 





The dispersion relation for Eqn. \ref{hamilt} is:
\begin{align} \label{disp_relation}
    E_{\pm}(k) = E_0 \pm \sqrt{(2C \cos{k} )^2 + \kappa^2}
\end{align}
where the positive and negative signs in Eqn. \ref{disp_relation} correspond to the upper and the lower bands, respectively. 
With our choice of the coupling constants in the $a$ and $b$ sublattices, the two bands are symmetric around $E_0$. 
%
The corresponding eigenstates are:  
\begin{align}\label{eigenstates}
    |\psi_{\pm}(k)\rangle = 
    \frac{1}{\sqrt{N_\pm(k)}} \begin{pmatrix}
    2C \cos{k} \pm \sqrt{(2C \cos{k})^2 + \kappa^2} \\ 
    \kappa
    \end{pmatrix}
\end{align}
where $|\psi_+ (k)\rangle$, $|\psi_- (k)\rangle$ correspond to the upper band $E_+(k)$ and the lower band $E_-(k)$, respectively. $N_\pm(k)$ is the corresponding normalization constant.

To implement the time boundary in our system, we impose a sudden change on some of the coupling constants in Eqn. \ref{hamilt}. As a numerical example, we consider a sudden change of $C$ in Eqn. \ref{hamilt} from $C = \kappa$ for $t < 0$ to $C = -4 \kappa$ for $t > 0$, creating a time boundary at $t = 0$. The band structures corresponding to $C = \kappa$ and $C = -4 \kappa$ are shown in Fig. \ref{fig1}b as blue and red lines, respectively. For each band structure, at a given $k$, there are two allowed eigenstates with different energies that propagate with opposite-signed group velocities of the same magnitude. Suppose at $t = -10/\kappa$, we excite one of the states with wavevector $k$, as indicated by the blue dot in Fig. \ref{fig1}b. For $t > 0$, after the time boundary, since the change is spatially uniform, the wavevector is conserved. We thus expect that the system should be in the two states as indicated by the red dots in Fig. \ref{fig1}b. These two states are the time-refracted wave that has the same sign in its group velocity as the initial state, and the time-reflected wave that has the opposite sign.


The effect of such a time boundary is demonstrated numerically in Fig. \ref{fig1}c, where we simulate $N=400$ total lattice sites, consisting of $M=200$ unit cells of $a$ and $b$ sites. At $t = -10/\kappa$, we initialize the system to be in the state
\begin{align}\label{gaussian_eigenstate}
    \begin{pmatrix}
    a_m \\ b_m
    \end{pmatrix} = \frac{1}{\sqrt{\sigma}\sqrt[4]{\pi}} e^{-(m-m_0)^2/2\sigma^2} e^{ik_0 m} | \psi_+ (k_0)\rangle
\end{align}
where $a_m$, $b_m$ are the amplitudes of the $a$ and $b$ sites in the $m$th unit cell, respectively. In Fig. \ref{fig1}c, we choose $k_0 = \pi/3 $, $m_0 = M/2$ and $\sigma = 5$. This state is a linear superposition of the eigenstates around the blue dot in Fig. \ref{fig1}b.  
As time evolves, the pulse initially propagates to the left in the $-n$ direction, consistent with the negative group velocity $v_g = dE/dk$ at the wavevector $k_0$ (blue dot in Fig. \ref{fig1}b).  At the time boundary at $t=0$, the pulse splits into two states which propagate with opposite-signed group velocities, in consistency with our discussion on the time-refracted and time-reflected waves.  

We now assume that both coupling constants $C$ and $\kappa$ can be varied at the time boundary.
In the case of an instantaneous change of the coupling constants, the time-refraction and time-reflection probabilities can be computed analytically. For an incident wave $|\psi_+ (k) \rangle$, which corresponds to the state in the upper band at momentum $k$ for the system before the time boundary, the probability of 
time refraction is:
\begin{align}\label{refrac_eqn}
T &= |\langle \widetilde{\psi}_+(k) | \psi_+(k) \rangle|^2 \nonumber \\
&= \frac{1}{N_+(k) \widetilde{N}_+(k)} \bigg[ \bigg(\mu(k) + \sqrt{\mu(k)^2 + \kappa^2} \bigg) \nonumber \\
&\hspace{2cm} \times \bigg(\widetilde{\mu}(k) + \sqrt{\widetilde{\mu}(k)^{2} + \widetilde{\kappa}^{2}} \bigg) + \kappa \widetilde{\kappa} \bigg]^2 
\end{align}
%
where tildes indicate parameters after the time boundary and $\mu(k) \equiv 2 C \cos{k} $. $|\widetilde{\psi}_-(k) \rangle$ corresponds to the eigenstate of the lower band $\widetilde{E}_-(k)$ at momentum $k$ after the time boundary.
The probability of time reflection is $R = 1-T$. The analogous probabilities for an incident state $|\psi_-(k)\rangle$ can be similarly computed.

To achieve complete reflection, we set the refraction probability $T = 0$, which requires that $\langle \widetilde{\psi}_+ | \psi_+ \rangle = 0$. In other words, the eigenstate of the upper band before the time boundary is orthogonal to the eigenstate of the upper band after the boundary.
%
%
Using Eqn. \ref{refrac_eqn}, we  obtain the condition for total time reflection: 
%
\begin{align}\label{complete_reflec_condition}
-\kappa \widetilde{\kappa} = \bigg(\mu + \sqrt{\mu^2 + \kappa^2}\bigg) 
\bigg(\widetilde{\mu} + \sqrt{\widetilde{\mu}^{2} + \widetilde{\kappa}^{2}} \bigg)
\end{align}
This condition is satisfied when
\begin{align}\label{complete_reflec_cond}
    H(t>0) - E_0 = - [H(t<0) - E_0]
\end{align}
Since the band structure of $H$ in Eqn. \ref{complete_reflec_cond} is symmetric with respect to $E_0$, the band structure remains the same before and after the time boundary. However, the eigenstate corresponding to the upper band before the boundary becomes that of the lower band after the boundary. In other words, $|\widetilde{\psi}_+(k) \rangle = |\psi_-(k)\rangle$ and $|\widetilde{\psi}_-(k)\rangle = |\psi_+(k)\rangle$. 

\begin{figure}
\includegraphics[width=0.48\textwidth]{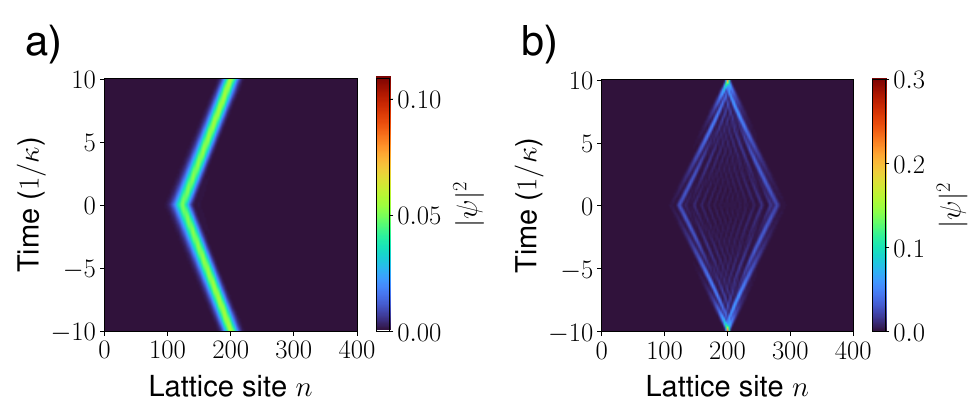}
\caption{\label{fig2} Numerical demonstration of complete time reflection. Shown here are the temporal evolutions of the state described by (a) Eqn. \ref{gaussian_eigenstate} and (b) Eqn. \ref{gaussian_no_eigenstate}. In both panels, the Hamiltonian parameters are $C = 5$ and $\kappa = 2$ before the time boundary. At the boundary $t=0$, the Hamiltonian instantaneously changes according to Eqn. \ref{complete_reflec_cond}.
}
\end{figure}

We confirm this effect of total time reflection numerically in Fig. \ref{fig2}a, where we initialize the system in the state described by Eqn. \ref{gaussian_eigenstate}. We use the same initialization parameters as in Fig. \ref{fig1}c. Before the time boundary $(t<0)$, the Hamiltonian has coupling constants $C = 5$ and $\kappa = 2$. After the time boundary ($t>0$), $C = -5$ and $\kappa = -2$. We notice the complete absence of the time-refracted wave after the time boundary. 

For the case when Eqn. \ref{complete_reflec_cond} is satisfied, the time boundary performs a time-reversal of the pulse envelope. To illustrate this, in Fig. \ref{fig2}b, we use the same time boundary setup as in Fig. \ref{fig2}a, but we initialize the system at $t = -10/\kappa$ with a state described by:
\begin{align}\label{gaussian_no_eigenstate}
\begin{pmatrix}
    a_m \\ b_m
    \end{pmatrix} = \frac{1}{\sqrt{2\sigma}\sqrt[4]{\pi}}e^{-(m-m_0)^2/2\sigma^2}  \begin{pmatrix}
    1 \\ 1
    \end{pmatrix}
\end{align}
where $m_0 = M/2$ and $\sigma =1$.
In contrast to the initial state for the case of Fig. \ref{fig2}a, here both bands are excited. Consequently, at $-10/\kappa < t < 0$, we see both forward and backward propagation in space.  
At the time boundary ($t = 0$), complete time reflection is achieved for every $k$ component. Consequently, after evolving the state for the same length of time after the time boundary ($t=10/ \kappa$), the system returns to the initial state.
%

The effect that we report here in Fig. \ref{fig2}b is reminiscent of Pendry's perfect lens based on negative index material \cite{negative_refrac_perfect_lens_pendry_PRL_2000}. However, in our system, we employ one time and one spatial dimension. 
A similar refocusing phenomenon at a source point has also been observed experimentally in water waves \cite{time_reversal_holography_fink_nature_physics_2016}. Our demonstration here differs in that we show complete time reflection for all wavevectors, whereas Ref. \cite{time_reversal_holography_fink_nature_physics_2016} shows both time reflection and time refraction. Time-reversal of the wave envelope has also been considered in Refs. \cite{time_revers_linear_optics_yanik_fan_PRL_2004, stopping_and_time_reversing_pulse_Sandhu_fan_OL_2007, all-linear_time_revers_chumak_nature_comm_2010}. None of these works \cite{time_revers_linear_optics_yanik_fan_PRL_2004, stopping_and_time_reversing_pulse_Sandhu_fan_OL_2007, all-linear_time_revers_chumak_nature_comm_2010}, however, employ a sharp time boundary as we present here. 

\begin{figure*}
\includegraphics[width=\textwidth]{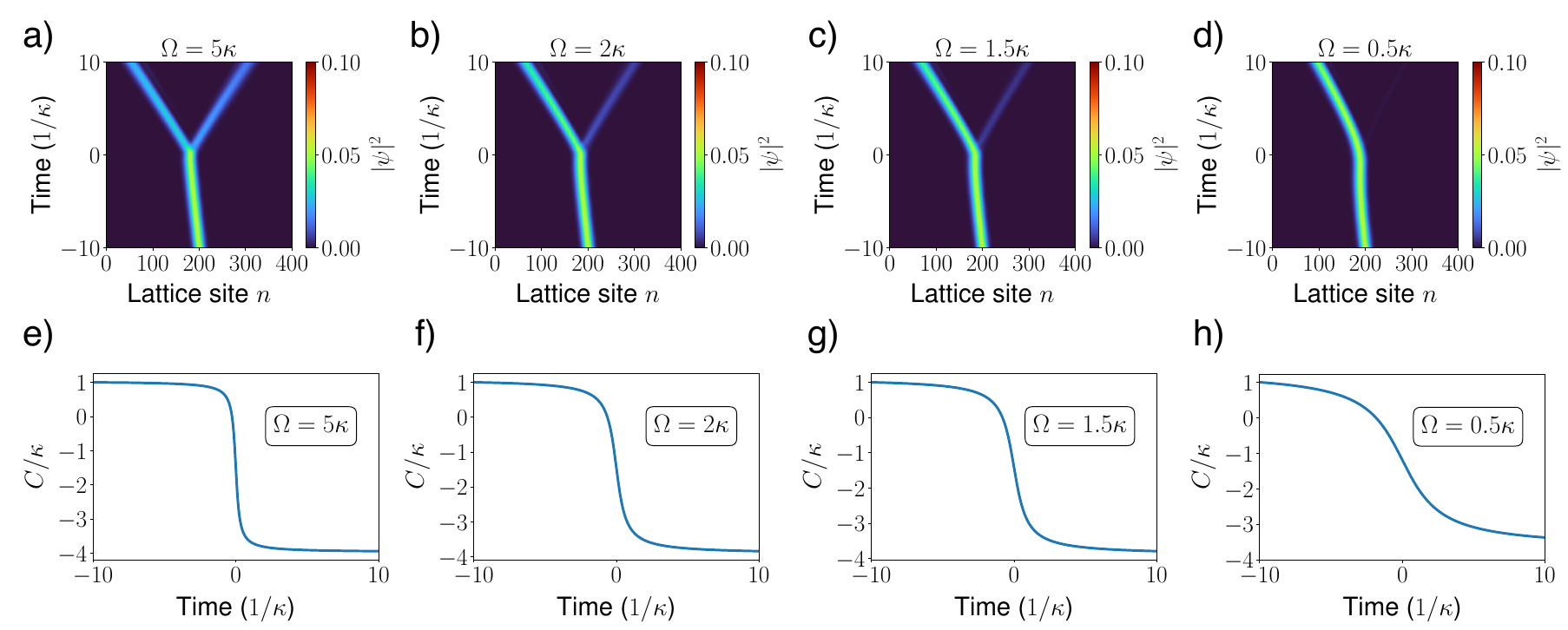}
\caption{\label{fig_smoothness} Numerical demonstration of smoothly varying time boundary. (a)--(d) Temporal evolution of wavefunction initialized in the state described by Eqn. \ref{gaussian_eigenstate}, using the same parameters as Fig. \ref{fig1}c. In the Hamiltonian, $C$ is varied smoothly from $C = \kappa$ to $C = -4 \kappa$, where $C(t)$ is given by Eqn. \ref{Ct_functional_form}. 
The band structures for $C = \kappa$ and $C = -4\kappa$ are shown in Fig. \ref{fig1}b. Simulations were performed using the 4th-order Runge-Kutta method. (e)--(h) Change in coupling constant $C$ as a function of time, depicted below its corresponding dynamical simulation. }
\end{figure*}

We now investigate the sharpness of the time boundary needed to observe time refraction and reflection. To facilitate the comparison, we use the same parameters as in Fig. \ref{fig1}c, but we vary $C(t)$ smoothly from $C = \kappa$ at $t = -10/\kappa $ to $C = -4 \kappa$ at $t = +\infty $ using the functional form: 
\begin{align}\label{Ct_functional_form}
    C(t) = - \frac{\delta}{\pi} \arctan{(\Omega t)} - \alpha(\Omega)
\end{align}
where $\delta = 5\kappa$ and $\alpha(\Omega)$ is a constant offset to ensure that $C(-10/\kappa) = \kappa$.
As can see in Fig. \ref{fig_smoothness}e--h, most of the variation of $C$ occurs around $t = 0$ within a time duration that is inversely related to $\Omega$. 
With Eqn. \ref{Ct_functional_form}, the sharpness of the time boundary is controlled by the parameter $\Omega$. 
Using different values of $\Omega$, we compare the temporal evolution of an initial pulse described by Eqn. \ref{gaussian_eigenstate} (Fig. \ref{fig_smoothness}a--d).
The corresponding change in the value of $C(t)$ is depicted in Fig. \ref{fig_smoothness}e--h.

When the time boundary is sharp ($\Omega = 5\kappa$), the incident pulse splits into time-reflected and time-refracted states (Fig. \ref{fig_smoothness}a). The resulting temporal evolution looks similar to that of the instantaneous time boundary (Fig. \ref{fig1}c).
As the frequency $\Omega$ decreases, we see that the time reflection effect becomes less prominent (Fig. \ref{fig_smoothness}b--d). At $\Omega = 0.5 \kappa$, the system is in the adiabatic regime where the time reflection is almost completely absent. Fig. \ref{fig_smoothness} indicates that the required switching rate for the time boundary in our system is on the order of $\kappa$ in order to achieve a significant time reflection effect. The required switching rate thus can be far lower than the frequency of the optical wave.

\begin{figure}
\includegraphics{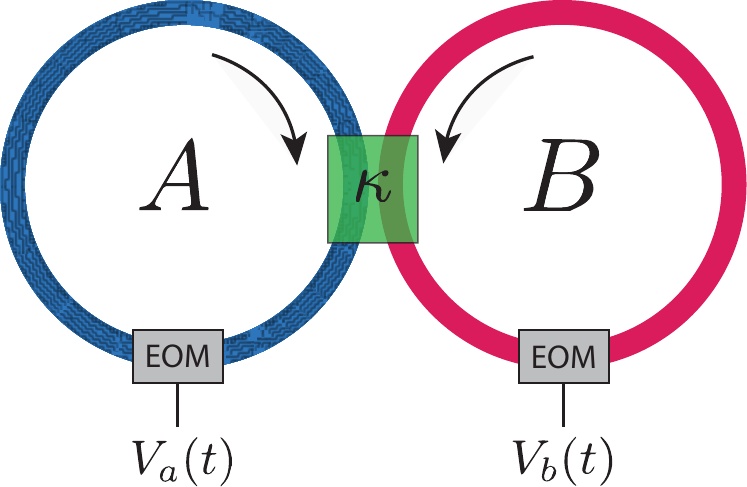}
\caption{\label{fig_ring_res} Synthetic frequency dimension implementation using dynamically modulated, coupled two-ring resonator system. $V_a(t)$ and $V_b(t)$ are the applied voltages, EOM is the electro-optical modulator, and $\kappa$ represents the directional coupler between the rings. Arrows indicate the counter-propagating modes needed to implement the two-band model.}
\end{figure}

The Hamiltonian in Eqn. \ref{hamilt} can in principle be implemented using coupled-resonator optical waveguides \cite{CROW_TB_model_Yariv_OL_99}, similar to what has been done in Ref. \cite{time_revers_linear_optics_yanik_fan_PRL_2004}. Below, however, we show that this Hamiltonian can be straightforwardly implemented using the concept of synthetic frequency dimension. 
Our proposed system consists of two coupled ring resonators, each modulated by an electro-optic modulator (EOM) driven by an applied voltage (Fig. \ref{fig_ring_res}). Each ring supports resonant frequency modes that differ by multiples of the free spectral range $\Omega_R=2\pi v_g/ L$, where $v_g$ is the group velocity and $L$ is the length of each ring. The frequency modes of ring A correspond to the $a$ lattice sites, while the modes of ring B are the $b$ sites in the two-leg ladder Hamiltonian (Fig. \ref{fig1}a). Since the lattice spacing has units of frequency in the synthetic frequency dimension, the momentum $k$ has units of time. The inter-leg coupling constant $\kappa$ is controlled by the splitting ratio of the directional coupler between the two rings. Due to the evanescent coupling between the rings, one needs to choose a pair of counter-propagating modes to implement the model. 

We assume that rings A and B are modulated with voltages $V_a(t) = V_0 \cos{(\Omega_R t)}$ and $V_b(t) = -V_0 \cos{(\Omega_R t )}$, respectively. Modulating each ring at the frequency of the FSR $\Omega_R$ results in the coupling between the nearest-neighbor frequency modes in each resonator \cite{luqi_synth_dim_photonics_Optica_2018, synth_freq_dynamic_mod_ring_res_luqi_avik_APL_2021,  avik_exp_band_struc_synth_dim_nature_2019}. The choice of the modulation waveforms, where the voltages are of the same modulation magnitude, but with opposite modulation phases, ensures that the coupling constants in each leg of the implemented Hamiltonian have equal magnitude and opposite signs. 

 

A detailed derivation showing that such a system can indeed implement the Hamiltonian of Eqn. \ref{hamilt} is in the Supplementary Material \cite{SM}. Here, we note that the switching rate ($\Omega$ in Eqn. \ref{Ct_functional_form}) needed to observe time reflection is on the order of the coupling constant $\kappa$, which can be in the microwave frequency regime if the rings are implemented using optical fibers. Experimental values of $\Omega_R = 2\pi \cdot 6$ MHz and $\kappa \approx 0.08$ $\Omega_R = 3$ MHz have been realized for such a system \cite{creating_bound_synth_freq_dim_avik_2022, kai_complex_energy_braiding_non-hermitian_bands_nature_2021}. 

In summary, we have shown that by utilizing a two-band model, time refraction and reflection of optical waves can be observed by imposing a temporal boundary on a time scale that is significantly slower than that of the carrier frequency.  
By using a system of coupled ring resonators, our model can be naturally implemented in synthetic frequency space. 
%
A natural extension would be to investigate a periodic array of such boundaries.


\begin{acknowledgments}
The authors thank Mohammed Benzaouia and Eran Lustig for helpful discussions. This work is supported by MURI projects from the U. S. Air Force Office of Scientific Research (Grant No. FA9550-18-1-0379, and FA9550-22-1-0339.)
O.L. acknowledges support from the NSF Graduate Research Fellowship and the Stanford Graduate Fellowship. 
\end{acknowledgments}



\bibliography{bibliography}

\clearpage
\appendix
\onecolumngrid
\section{Supplementary Material}
\subsection{Derivation of coupled mode equations}\label{sec_deriv_coupled_mode_eqns}

In this section, we derive the coupled mode equations governing the coupled two-ring resonator system \cite{avik_HOTI_SSH_2ring_resonator_2020, siddarth_arbitrary_linear_trans_synth_dim_nature_2021, luqi_synth_dim_photonics_Optica_2018}.

Assuming nonmagnetic material ($B = \mu_0 H$), we start from Maxwell's equations:
\begin{equation}\label{maxwell_eqn}
    \curl \curl \mathbf{E} = -\mu_0 \frac{\partial^2}{\partial t^2} (\epsilon \mathbf{E})
\end{equation}

We first consider Ring A. Assuming the refractive index is modulated perturbatively by the electro-optic modulator in the ring resonator, $\epsilon(t) = \epsilon_{s} + \delta \epsilon(t)$ where $\epsilon_{s}$ is the static value of the the permittivity. The free spectral range (FSR) of the unmodulated ring with permittivity $\epsilon_s$ is $\Omega=2\pi v_g/ L=2\pi c/L\sqrt{\epsilon_s}$, where $v_g$ is the group velocity and $L$ is the length of the ring. Using the modulated permittivity $\epsilon(t)$ in Eqn. \ref{maxwell_eqn}, we have:
\begin{equation} \label{expanded_maxwell}
    \curl \curl \mathbf{E} = -\mu_0 \epsilon_{s} \frac{\partial^2 \mathbf{E}}{\partial t^2} - \mu_0 \frac{\partial^2}{\partial t^2}[\delta \epsilon(t) \mathbf{E}] 
\end{equation}

For Ring A, in the absence of modulation or coupling to Ring B, we can expand the electric field in terms of its frequency components:
\begin{equation}\label{elec_field_expansion}
    \mathbf{E}(\mathbf{r},t) = \sum_m a_m(t) \mathbf{E}_m(\mathbf{r}) e^{i\omega_m t} 
\end{equation}
%
where $a_m(t)$ is the time-dependent, complex amplitude and $\mathbf{E}_m(\mathbf{r})$ is the spatial profile of the $m$th frequency eigenmode of the unperturbed system (no modulation) at frequency $\omega_m = \omega_0 + m\Omega$, where $\omega_0$ is the central carrier wave frequency. 


Plugging Eqn. \ref{elec_field_expansion} into Eqn. \ref{expanded_maxwell}, we have:
\begin{align}\label{sub_in_E_fourier}
    \sum_m a_m(t) e^{i\omega_m t} &[ \curl \curl \mathbf{E}_m(\mathbf{r}) ] = - \mu_0 \epsilon_s \sum_m \mathbf{E}_m(\mathbf{r}) \frac{\partial^2}{\partial t^2}[a_m(t) e^{i\omega_m t}] \nonumber  \\
    & \hspace{10em} 
    - \mu_0 \sum_n \mathbf{E}_n(\mathbf{r}) \frac{\partial^2}{\partial t^2}[ \delta \epsilon(t) a_n(t) e^{i\omega_n t} ] \nonumber \\ 
    &= 
        - \mu_0 \epsilon_s \sum_m \mathbf{E}_m(\mathbf{r}) \bigg[ \bigg( \frac{d^2 a_m}{dt^2} + 2i\omega_m \frac{da_m}{dt} - \omega_m^2 a_m \bigg) e^{i\omega_m t} \bigg]  \nonumber \\
    & \hspace{10em} - \mu_0 \sum_n \mathbf{E}_n(\mathbf{r}) \frac{\partial^2}{\partial t^2}[ \delta \epsilon(t) a_n e^{i\omega_n t} ] 
\end{align}
%
We know that $\curl \curl \mathbf{E}_m(\mathbf{r}) e^{i\omega_m t} =  \mu_0 \epsilon_s \omega_m^2 \mathbf{E}_m(\mathbf{r}) e^{i\omega_m t} $
in the unperturbed system.
We also ignore the $\frac{d^2 a_m}{dt^2}$ terms since we assume the slowly varying envelope approximation. Eqn. \ref{sub_in_E_fourier} now becomes:
\begin{align}\label{simplified_eqn_4}
    0 &= 
    - \mu_0 \epsilon_s \sum_m \mathbf{E}_m(\mathbf{r})  \bigg( 2i\omega_m \frac{da_m}{dt} \bigg) a_m e^{i\omega_m t}   
    - \mu_0 \sum_n \mathbf{E}_n(\mathbf{r}) \frac{\partial^2}{\partial t^2}[ \delta \epsilon(t) a_n e^{i\omega_n t} ] 
\end{align}
If we apply an electro-optic modulation at the frequency of one FSR ($\Omega$) with a phase $\gamma$, $\delta \epsilon (t)$ is:
\begin{equation}
\delta \epsilon (t)= \delta (\mathbf{r}) \cos{(\Omega t + \gamma)} = \frac{\delta(\mathbf{r})}{2} \bigg(e^{i (\Omega t + \gamma)} + e^{-i (\Omega t + \gamma)} \bigg)
\end{equation}
where $\delta (\mathbf{r})$ includes the modulation strength and spatial distribution of the modulation. 
Eqn. \ref{simplified_eqn_4} becomes:
\begin{align}\label{eqn5}
    \sum_m \mu_0 \epsilon_s \mathbf{E}_m(\mathbf{r}) \bigg( 2i\omega_m \frac{da_m}{dt} \bigg) e^{i\omega_m t} &=
     - \sum_n \mu_0 \mathbf{E}_n(\mathbf{r}) \frac{\delta(\mathbf{r})}{2} \frac{d^2}{dt^2} \bigg[ a_n(t) e^{i\omega_{n+1} t} e^{i \gamma} + a_n(t) e^{i\omega_{n-1} t} e^{-i \gamma} \bigg]  \nonumber \\ 
    &  \approx  - \sum_n \mu_0 \mathbf{E}_n(\mathbf{r}) \frac{\delta(\mathbf{r})}{2} \bigg[ a_n (-\omega_{n+1}^2 e^{i \omega_{n+1} t} e^{i\gamma} - \omega_{n-1}^2 e^{i \omega_{n-1} t} e^{-i\gamma} ) \bigg]
\end{align}
where we have ignored $\frac{d^2 a_n}{dt^2}$ and $\frac{da_n}{dt}$ terms on the RHS due to the slowly varying envelope approximation and the weak modulation strength.

Matching the terms with coefficient $e^{i\omega_m t}$ on both sides of Eqn. \ref{eqn5}, we get:
\begin{align}
    \epsilon_s \mathbf{E}_m(\mathbf{r}) 2i\omega_m \frac{da_m}{dt} &=    \frac{\delta(\mathbf{r})}{2} \omega_m^2 \bigg(a_{m-1} \mathbf{E}_{m-1}(\mathbf{r})   e^{i\gamma} + a_{m+1} \mathbf{E}_{m+1}(\mathbf{r})  e^{-i\gamma} \bigg) 
\end{align}

Taking the dot product with $\mathbf{E}_{m}^*$, integrating over space, and using the normalization $\int \mathbf{E}_{m}^* \cdot \epsilon_s \mathbf{E}_{m} = 1$, we have:
%
\begin{align}
      2i \omega_m \frac{da_m}{dt} &=   \frac{\omega_m^2}{2} \bigg[ a_{m-1} e^{i\gamma} \int (\mathbf{E}_{m}^* \cdot \delta(\mathbf{r}) \mathbf{E}_{m-1}) d\mathbf{r} + a_{m+1}e^{-i\gamma} \int (\mathbf{E}_{m}^* \cdot \delta(\mathbf{r}) \mathbf{E}_{m+1}) d\mathbf{r}
    \bigg] 
\end{align}
Dividing both sides by $2\omega_m$, we get:
\begin{align}
    i \frac{da_m}{dt} &=  \frac{\omega_m}{4} \bigg[ a_{m-1} e^{i\gamma} \int (\mathbf{E}_{m}^* \cdot \delta(\mathbf{r}) \mathbf{E}_{m-1}) d\mathbf{r} + a_{m+1}e^{-i\gamma} \int (\mathbf{E}_{m}^* \cdot \delta(\mathbf{r}) \mathbf{E}_{m+1}) d\mathbf{r}
    \bigg] \nonumber \\
    &\equiv C_{ m-1} a_{m-1} +  C_{m+1}a_{m+1}
\end{align}
where we define:
\begin{equation}\label{Ca_eqn1}
C_{m-1} \equiv  \frac{\omega_m}{4} e^{i\gamma} \int (\mathbf{E}_{m}^* \cdot \delta(\mathbf{r}) \mathbf{E}_{m-1}) d\mathbf{r}
\end{equation}

\begin{equation}\label{Ca_eqn2}
C_{m+1} \equiv  \frac{\omega_m}{4} e^{-i\gamma} \int (\mathbf{E}_{m}^* \cdot \delta(\mathbf{r}) \mathbf{E}_{m+1}) d\mathbf{r}
\end{equation}
%
For a ring resonator based on a fiber with a single spatial mode and a single fixed polarization mode, it is reasonable to assume that $\mathbf{E}_m$ does not depend on $m$. Assuming only the real part of the refractive index is modulated (i.e. $\delta (\mathbf{r})$ is real), $C_{m-1} = C_{m+1}^*$. Therefore, for Ring A we can define a single coupling constant without $m$ dependency: $C_a \equiv C_{m-1}=C_{m+1}^*$. Moreover, while in general $C_a$ is complex-valued, by choosing the modulation phase $\gamma=0$ or $\pi$ we can obtain a real positive or real negative $C_a$, respectively. 

%

Since the $m$th mode of Ring A is coupled to the $m$th mode of Ring B via a directional coupler at a constant rate $\kappa \in \mathbb{R}$ that does not depend on $m$, we add an extra term describing such an inter-ring coupling and obtain the final equation of motion for $a_m$: 
\begin{equation}\label{a_m_eqn_of_motion}
    i\frac{da_m}{dt} =  C_a a_{m-1} + C_a^* a_{m+1} + \kappa b_m
\end{equation}
where $b_m$ is the complex amplitude of the $m$th frequency eigenmode in Ring B.  
%
Similarly, the modal amplitudes $b_m$ in Ring B obey the equation of motion:
\begin{equation}
    i \frac{db_m}{dt} = C_b b_{m-1} +C_b^* b_{m+1} + \kappa a_m
\end{equation}

%
Writing the two equations of motion together in matrix form:
\begin{equation}
    i \begin{pmatrix}
\dot{a}_m \\ \dot{b}_m
\end{pmatrix} = \begin{pmatrix}
C_a & 0 \\
0 & C_b
\end{pmatrix} \begin{pmatrix}
a_{m-1} \\ b_{m-1}
\end{pmatrix}
+ \begin{pmatrix}
C_a^* & 0 \\
0 & C_b^*
\end{pmatrix} \begin{pmatrix}
a_{m+1} \\ b_{m+1}
\end{pmatrix}
+
\begin{pmatrix}
0 & \kappa \\
\kappa & 0
\end{pmatrix}
\begin{pmatrix}
a_m \\ b_m
\end{pmatrix}
\end{equation}
%

If we set $C_a = -C_b \equiv C \in \mathbb{R}$ (e.g. by setting $\gamma = 0$ in Ring A and $\gamma = \pm \pi$ in Ring B), we achieve the tight-binding Hamiltonian: 
\begin{align} \label{hamilt_supp}
    H =  \sum_m [C \hat{a}_m^\dagger \hat{a}_{m+1} + \text{h.c.}] 
    + \sum_m [-C \hat{b}_m^\dagger \hat{b}_{m+1} + \text{h.c.}] 
    &+ \sum_m [\kappa \hat{a}_m^\dagger \hat{b}_m + \text{h.c.}] 
\end{align}
where 
$m$ is the lattice site index, which corresponds to the $m$th frequency mode in synthetic space. Since $\hat{a}_m, \hat{b}_m$ describe the slowly varying envelopes of the $m$th frequency mode $e^{i\omega_m t}$, the band structure of Eqn. \ref{hamilt_supp} is centered around a central reference energy $m\Omega$ and repeats every $\Omega$. Thus, $E_0$ in Eqn. \ref{hamilt} of the main text corresponds to $m\Omega$ in our synthetic frequency implementation. 
%
%
%
\subsection{Proposed experimental setup}
To observe the evolution of an initial state subject to a time boundary, we propose to use the coupled two-ring resonator system (Fig. 4 in main text). To prepare an excitation around a certain $k$ value, we can use an amplitude modulator to produce a single pulse from a continuous-wave (CW) laser. Since time is inherently $k$ in this system and one round-trip time corresponds to one Brillouin zone, 
a certain $k$ can be excited by controlling the pulse's center within one round-trip time. 
Depending on the scale of the free spectral range, the detection can either be based on an optical spectrum analyzer that directly measures the spectrum, or be based on fast time-resolved intensity measurements.
For time-resolved measurements, a heterodyning technique can be used to obtain phase information \cite{heterodyning_paper_dutt_experimental_2019}. 
The beat pattern in time formed by the interference of the frequency-shifted CW laser with the output signals can give the time-dependent phase information from the time-resolved intensity measurements. By Fourier transforming such heterodyne signals, it is possible to reconstruct the spectral components of each frequency mode at $\omega_m=\omega_0+m\Omega$. These measurements can be taken over time to map out the evolution of the state in synthetic frequency space. After the time boundary is implemented, the time-reflected pulse can be identified by detecting a frequency shift in the opposite direction as that of the incident pulse. A frequency shift in the same direction as the incident pulse can be attributed to the time-refracted pulse. 

\end{document}